\documentclass[runningheads]{llncs}
\usepackage{graphicx}
\usepackage{float}
\usepackage{amsmath}
\usepackage{amssymb}

\usepackage{caption}
\usepackage{subcaption}

\usepackage{hyperref}
\usepackage{xcolor}

\begin{document}
\title{Adversarial Continual Learning for Multi-Domain Hippocampal Segmentation\protect\footnote{Supported by the Bundesministerium f{\"u}r Gesundheit (BMG) with grant [ZMVI1- 2520DAT03A]. The final version of this manuscript will be published in Springer Lecture Notes in Computer Science, Domain Adaptation and Representation Transfer at Medical Image Computing and Computer Assisted Intervention - DART at MICCAI 2021 as part of the joint MICCAI Workshops proceedings (doi will follow).}}
\titlerunning{Adversarial Continual Learning for Multi-Domain Segmentation}

\author{Marius Memmel\inst{1} \and
Camila Gonzalez\inst{1} \and
Anirban Mukhopadhyay\inst{1}}

\authorrunning{M. Memmel et al.}
\institute{$^1$Technical University of Darmstadt, Karolinenpl. 5, 64289 Darmstadt, Germany}

\maketitle              
\begin{abstract}
Deep learning for medical imaging suffers from temporal and privacy-related restrictions on data availability. To still obtain viable models, continual learning aims to train in sequential order, as and when data is available. The main challenge that continual learning methods face is to prevent catastrophic forgetting, i.e., a decrease in performance on the data encountered earlier. This issue makes continuous training of segmentation models for medical applications extremely difficult. Yet, often, data from at least two different domains is available which we can exploit to train the model in a way that it disregards domain-specific information. We propose an architecture that leverages the simultaneous availability of two or more datasets to learn a disentanglement between the content and domain in an adversarial fashion. The domain-invariant content representation then lays the base for continual semantic segmentation. Our approach takes inspiration from domain adaptation and combines it with continual learning for hippocampal segmentation in brain MRI. We showcase that our method reduces catastrophic forgetting and outperforms state-of-the-art continual learning methods.

\keywords{continual learning \and adversarial training \and feature disentanglement \and hippocampus MRI segmentation}
\end{abstract}

\section{Introduction}
In medical imaging, privacy regulations and temporal restrictions limit access to data \cite{Gonzalez2020}.
These limitations inhibit the application of traditional supervised deep learning methods for medical imaging tasks, which require the simultaneous availability of all data during training.
Continual learning reframes the problem into a sequential training process, where not all datasets are available at each time step. However, when we evaluate continual learning models, they still experience a significant drop in performance, caused by \textit{catastrophic forgetting}, i.e., the model adapting too strongly to particularities of the last training batch \cite{Kirkpatrick2015}.

The leading cause of catastrophic forgetting in medical imaging is multi-domain data originating from different domains \cite{Pianykh2020}. These domains result from diverse disease patterns among the examined subjects and divergent technologies and standards used during the acquisition process.
In magnetic resonance imaging (MRI) it is common practice for institutions to operate scanners from various vendors and employ disparate protocols \cite{Li2020}. Additionally, MRI datasets frequently contain subjects that are either healthy or suffer from various pathological conditions.
To sufficiently solve a task with data from multiple domains, models have to adapt and learn in a domain-invariant fashion.

\begin{figure}[t]
    \begin{subfigure}{.95\textwidth}
         \centering
         \includegraphics[width=.95\textwidth]{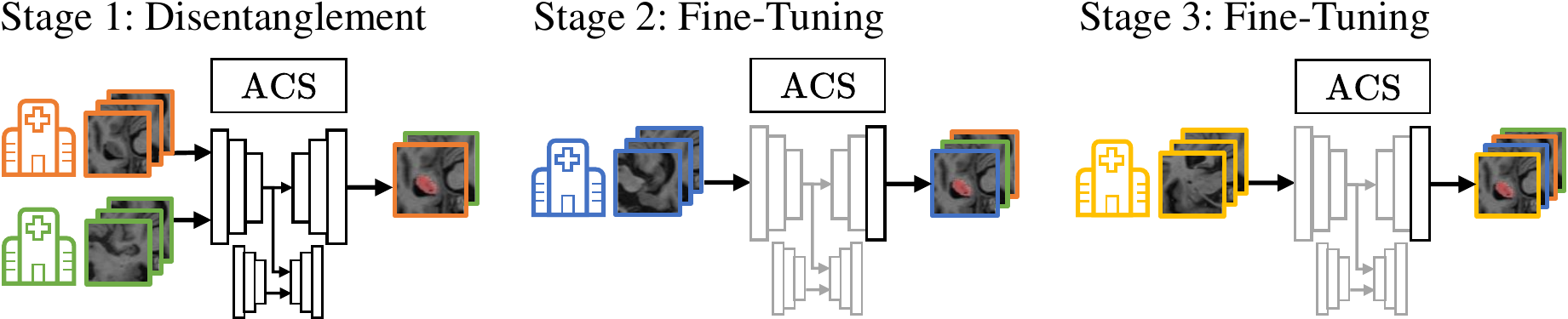}
         \caption{}
         \label{fig:scenario_1}
     \end{subfigure}
     \hfill
     \begin{subfigure}{.95\textwidth}
         \centering
         \includegraphics[width=.95\textwidth]{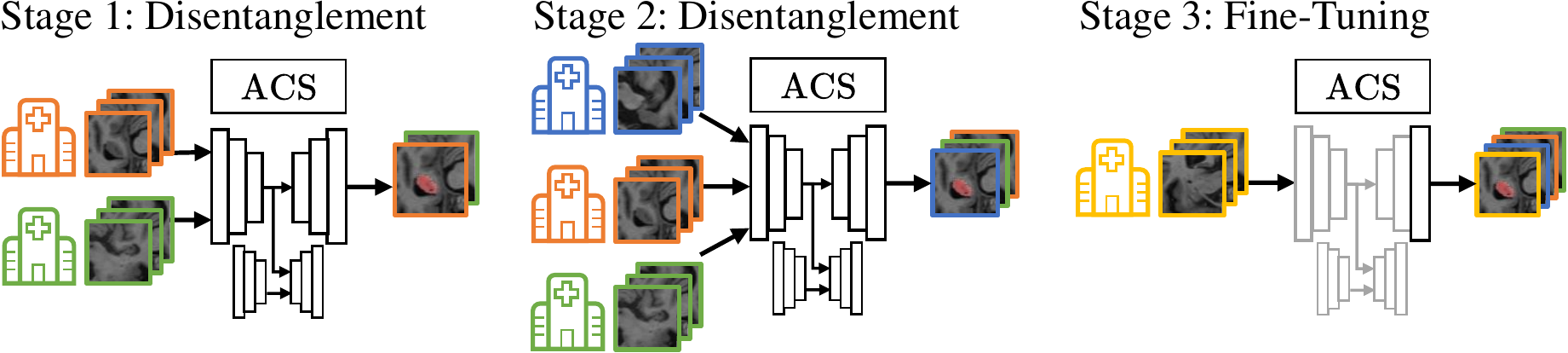}
         \caption{}
         \label{fig:scenario_2}
     \end{subfigure}
    \caption{The Adversarial Continual Segmenter (ACS) can react to different dataset availability through training (black) or freezing (gray) network parts.
    a) In stage 1, the disentanglement is trained on two datasets from different domains. In stage 2, a new dataset is available, and the previous ones are not. ACS is now fine-tuned on the new one. As a fourth dataset is introduced in stage 3, fine-tuning is repeated. b) If the first two datasets are still available when a third becomes accessible, disentanglement can be repeated in stage 2 with three datasets.} 
    \label{fig:overview}
\end{figure}

The limited availability of multi-domain data makes developing a general-purpose model for continual hippocampal segmentation difficult.
Commonly, at least two datasets from different domains are accessible simultaneously due to open access, relaxed restrictions, or access to historical data acquired with older scanners and protocols within the institution.
This serendipity allows for learning a disentanglement between the domains and the content needed for segmentation, which continual learning methods do not yet exploit.
Inspired by image-to-image translation (I2I), we utilize adversarial training to learn a disentanglement between a domain-invariant content representation sufficient for segmentation and a dataset-specific domain representation \cite{Huang,Kamnitsas2017,Lee2018}.
We train an encoder for each representation and share the content encoder with our segmentation module.
Finally, we extend our approach to continual learning. In Fig. \ref{fig:overview}, we describe how our architecture could react to common dataset availability scenarios. We perform experiments on a subset of those hypothetical scenarios.

\textbf{We contribute the \textit{Adversarial Continual Segmenter} (ACS)} for continual semantic segmentation of multi-domain data through adversarial disentanglement and latent space regularization that reduces catastrophic forgetting in hippocampal segmentation of brain MRIs.

\section{Related work}
\textbf{Adversarial disentanglement:} Several Generative Adversarial Networks (GANs) disentangle the feature space to improve interpretability \cite{Kazeminia2020}. 
Chen et al. \cite{Chen2016} take a mutual-information-based approach while Karras et al. \cite{Karras2019} directly modify the generator to achieve automatic separation of high-level attributes.
Adversarial disentanglement shows promising results when applied to segmentation in a multi-domain \cite{Jiang2020} and multi-modal setting \cite{Chartsias2019}.
In domain adaptation, Kamnitsas et al. \cite{Kamnitsas2017} utilize domain-invariant features for a segmentation task \cite{Toeffi} and learn those with adversarial regularization of numerous layer outputs.

\textbf{Cross-domain disentanglement:} I2I translation extends cross-domain feature disentanglement by splitting the latent space into a content and style encoding to achieve better translation results \cite{Liu}. The content encoding is assumed only to capture task-specific information. The style encoding holds the domain-specific information. Huang et al. \cite{Huang} and Lee et al. \cite{Lee2018} further assume that the complexity of the content outweighs the domain, which they reflect in different encoder complexities.

\textbf{Continual learning:} 
The main problem that continual learning methods face is catastrophic forgetting.
For this purpose, regularization-based approaches constrain important parameters from changing \cite{Baweja2018,VanGarderen2019,Kirkpatrick2015,Lenga2020}.
With a similar goal, Memory Aware Synapses (MAS) \cite{Aljundi2018} learn importance weights for each parameter and use those to penalize parameter changes. 
Knowledge distillation methods try to preserve specific model outputs to retain performance on old data \cite{Douillard2020,Michieli2019a}.
Keeping a subset of training data is also widely used, e.g., in dynamic memory \cite{Hofmanninger2020} and rehearsal methods \cite{Sokar2021}.
However, keeping parts of the old data is not feasible in most medical imaging scenarios due to privacy concerns \cite{Gonzalez2020}.

Whereas existing continual learning methods focus solely on a sequential learning process and do not consider the simultaneous availability of datasets and their divergent domains, we specifically exploit these circumstances through adversarial disentanglement.

\begin{figure}[tb]
    \centering
    \includegraphics[width=1\textwidth]{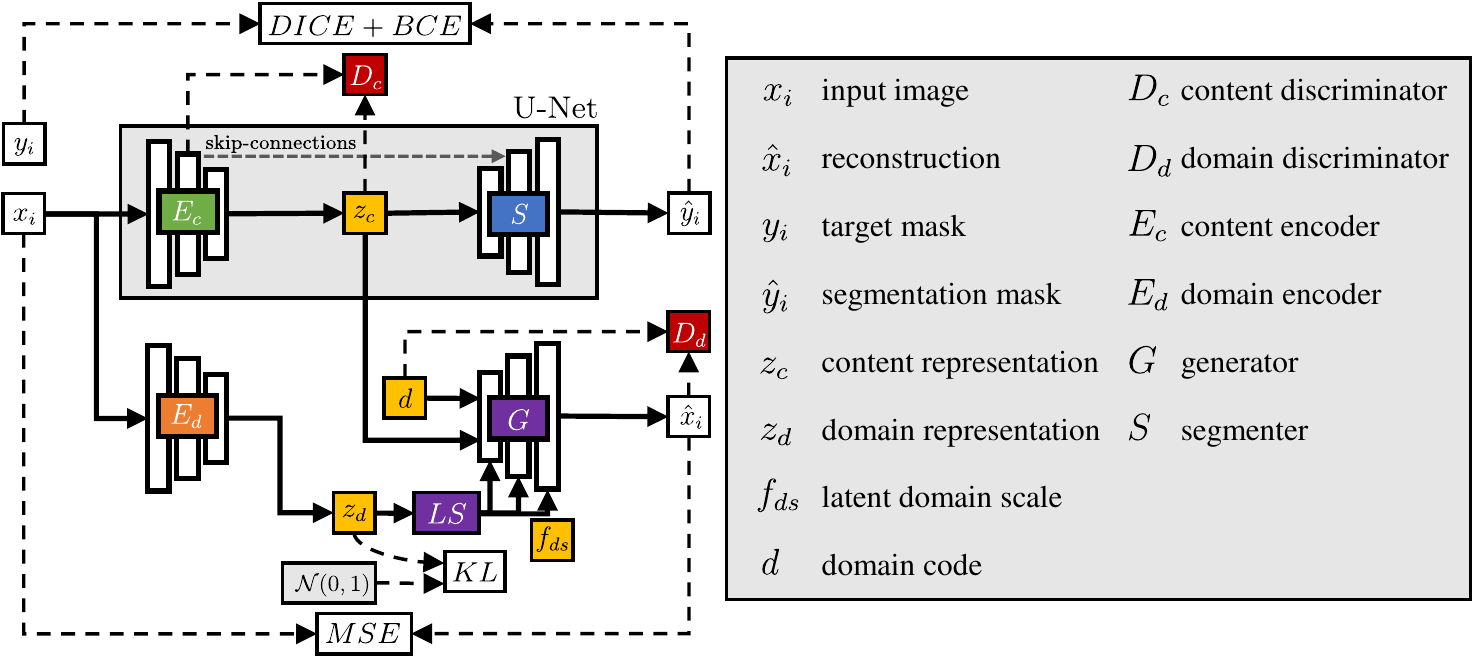}
    \caption{Detailed visualization of the ACS architecture. We achieve feature disentanglement through two encoders, $E_c$ for content and $E_d$ for domain. The content and domain discriminators $D_c$ and $D_d$ regularize the latent representation $z_c$, and in the case of $D_c$, also the skip-connections of the U-Net to be domain-invariant. Segmentation is done by a forward pass of the U-Net.}
    \label{fig:architecture}
\end{figure}

\section{Methods}
\label{sec:method}
We first describe how we disentangle input images into content and domain representation and follow up by introducing the adversarial approach. Our domain representation models the heterogeneity of the acquisition modality, e.g., varying protocols and machine vendors, as well as different disease patterns.
To learn the domain-invariant content representation, we initially train on two datasets simultaneously, as is common in I2I translation.
This representation then acts as a basis for our U-Net segmenter.

\label{subsec:feature_disentanglement}
\textbf{Variational autoencoder loss:}
We model the domain encoder as a variational autoencoder (VAE) \cite{kingma2014}, which encodes the domain as a distribution $z_d$ parameterized by variance $\sigma^2$ and mean $\mu$. Because the complexity of the domain is assumed to be lower than the content complexity, we limit the dimensionality of the domain representation to one float value. We use a combination of a reconstruction loss between the input image $x_i$ and the generator output $\hat{x}_i$ and a Kullback–Leibler regularization term weighted by hyperparameter $\eta$ to draw the encoding close to a normally distributed Gaussian prior of $\mathcal{N}(0,1)$.

\begin{equation}
    \mathcal{L}_{VAE} = \sum_{i} \|\hat{x}_i - x_{i}\|^2_2 +\eta \mathop{\mathbb{E}_{x_i \sim X_i, z \sim \mathcal{N}(0,1)}} \left[ KL\left(E_d\left(x_i \right )||z\right )\right ]
\end{equation}

\textbf{Latent regression loss:}
To prepare the domain information for the generator input, we first sample from the learned domain distribution and then pass the samples into the latent scale layer $LS$ as proposed by Alharbi et al. \cite{Alharbi2019} to produce a latent domain scale $f_{ds}$. 
To give $G$ additional information about the domain, we inject a domain code $d$ using central biasing instance normalization \cite{Jiang2020,Yu2018}. We model this domain code with a one-hot vector.
As proposed by Jiang and Veeraraghavan \cite{Jiang2020} and based on Lee et al. \cite{Lee2018} and Huang et al. \cite{Huang}, we also define a latent code regression loss $\mathcal{L}_{lr}$, which constrains the generator to produce unique mappings for a latent code $z$. 


\textbf{GAN loss:}
We interpret the encoder of the U-Net as content encoder $E_c$ with the output of the bottom layer treated as content representation $z_c$.
Both $z_c$ and the scaled domain sample $f_{ds}$ are fed into generator $G$ to reconstruct the input sample $x_i$.
To now disentangle the content and domain, we introduce adversarial training.
We deploy a domain discriminator $D_d$ that regularizes $E_c$, $E_d$, and $G$ by discriminating whether an input image $x_i$ is part of a given domain $d$. $D_d$ trains on a combination of real and generated images as described in the corresponding discriminator loss in Eq. \ref{eq:L_GAN_D}. $G$ generates these images from content $z_c$ of $x_i$ and random domain $z$.
$G$, $E_c$, and $E_d$ counter the discriminator by minimizing a negative binary cross-entropy loss shown in Eq. \ref{eq:L_GAN_G}.
The training forces $E_c$ to produce a domain-invariant output, which we utilize for the segmentation task.

\begin{equation}\label{eq:L_GAN_D}
\begin{aligned}
    \mathcal{L}_{GAN_D} = \sum_{i} & \mathop{\mathbb{E}_{x_{i}\sim X_{i}}}\left[\log(D_{d}\left(x_i, d_i \right )\right ]\\ 
        & + \mathop{\mathbb{E}_{x_{i}\sim X_{i}}}\left[\log(1-D_{d}\left(G\left(E_{c}\left(x_i \right ), LS\left(z \right ) \right ), d_{i} \right )\right]
\end{aligned}
\end{equation}
\begin{equation}\label{eq:L_GAN_G}
\mathcal{L}_{GAN_G} = -\sum_{i} \mathop{\mathbb{E}_{x_{i}\sim X_{i}, z \sim \mathcal{N}(0,1)}}\left[\log(D_{d}\left(G\left(E_{c}\left(x_i \right ), LS\left(z \right ) \right ), 1 \right )\right]
\end{equation}

\textbf{Content adversarial loss:}
The information about the domain can still flow from $E_c$ to $S$ via the skip-connections of the U-Net. To prevent this, we introduce a content discriminator $D_c$ inspired by Jiang et al. and Kamnitsas et al. \cite{Jiang2020,Kamnitsas2017}.
$D_c$ regularizes $z_c$ as well as the skip-connections because we want $E_c$ to not leak any domain information to the segmenter $S$.
The discriminator design is similar to a reversed U-Net decoder, i.e., it takes $z_c$ as input and the skip-connections of $E_c$ at each corresponding layer.
We train both $D_c$ and $E_c$ using a multi-class cross-entropy loss $\mathcal{L}_{adv_{D_{c}}}^c$ and $\mathcal{L}_{adv_{E_{c}}}^c$ respectively. In the case of the generated images, we represent the domain code as a placeholder class. To ensure that the adversarial training is stable, we train the discriminators and the remaining architecture at separate steps \cite{Jiang2020}.



\textbf{Segmentation:} \label{subsec:segmentation}
To produce the segmentation mask $\hat{y}_i$, we encode input $x_i$ into $z_c$ through $E_c$. We then pass $z_c$ and the skip-connections of $E_c$ into $S$ to compute $\hat{y}_i$. We train the U-Net for semantic segmentation through a pixel-wise combination of a Dice and a binary cross-entropy loss between the target mask $y_i$ and the prediction $\hat{y}_i$. 
After initially training the architecture on two or more datasets, the model has sufficiently learned the disentanglement between the content and the domain. To train on a new dataset, we only fine-tune the last four convolutional layers of $S$.

\section{Datasets \& Experiments}
\label{sec:experiments}
\textbf{Datasets:} All datasets are from different domains and contain T1-weighted MRIs.
The first dataset was released as part of the \textit{2018 Medical Segmentation Decathlon} challenge \cite{Decathlon} and consists of 195 subjects in total, with 90 healthy and 105 with non-affective psychotic disorder.
The scans were collected using a Philips Achieva scanner, and the mean size of the volumes is $35\times49\times35$.
The second dataset was published in \textit{Scientific Data} \cite{Dryad} and has a T1-weighted dataset with 25 healthy subjects. All scans were acquired using MRI systems with 3 Tesla units. The mean standard resolution is $48\times64\times64$.
Finally, the third dataset is provided by the \textit{Alzheimer's Disease Neuroimaging Initiative} \cite{HarP} and consists of 68 subjects that are either part of the control group or suffer from either mild cognitive impairment or Alzheimer's disease. The images were acquired with scanners from Siemens, GE, and Philips with 23, 24, and 21 scans, respectively. The mean volume size is $48\times64\times64$.
All three datasets provide reference segmentation masks for the hippocampus. The masks were annotated manually with the protocols defined in the respective publications.
We evaluate our architecture on all three datasets, which we will refer to as A, B, and C, respectively.

\textbf{Experimental setup:}
We split each dataset into 70\% train, 20\% test, and 10\% validation and use the latter to select the hyperparameters. We train slice-by-slice and upsample via bilinear interpolation to achieve uniform slices.
We compare ACS with the following baselines. First, just the U-Net block of ACS (U-Net-b) shown in Fig. \ref{fig:architecture}, and second a standard U-Net.
Furthermore, we extend the U-Net by knowledge distillation on the output layer (OL-KD) as proposed by Michieli and Zanuttigh \cite{Michieli2019a}, and Memory Aware Synapses adapted to brain segmentation (BS-MAS) by Özgün et al. \cite{Ozgun2020}. As suggested in BS-MAS, we divide the surrogate loss by the number of network parameters and normalize the resulting importance values between zero and one. We report the Intersection over Union (IoU) and Dice coefficient on the hippocampus class of the test set.
We use a batch size of 40 and train on four Tesla V100 SXM3 GPUs.
Each method receives training over 60 epochs. After 30 epochs, the training only continues with the third dataset.
We repeat training for every combination of the three datasets (\textit{AB-C}, \textit{AC-B}, \textit{BC-A}), e.g., initial training on datasets A and B, then on C (\textit{AB-C}). Additionally, we jointly train ACS and the U-Net on all datasets simultaneously (\textit{ABC}). To justify the necessity for all mechanisms in our method, as described in Sec. \ref{sec:method}, we conduct an ablation study in Tab. \ref{tab:ablation}.
Implementation details and qualitative results including the disentanglement can be found in the supplementary material and code on {\color{blue}\href{https://github.com/MECLabTUDA/ACS}{github.com/MECLabTUDA/ACS}}.

\section{Result \& Discussion}
To assess the continual learning performance, we evaluate the results after stages 1 and 2 corresponding to Fig. \ref{fig:scenario_1}. An ideal algorithm should perform equally or better on the initial training datasets from epoch 30 to 60 while it should improve on the third dataset added after 30 epochs.

\textbf{Stage 1:} Tab. \ref{tab:results_30} shows the results for all methods after 30 epochs on the initial two datasets. All baselines observe the same score because they apply the regularization in the second training stage, whereas ACS performs disentanglement during the initial training phase. ACS outperforms them by a Dice of $+0.178\pm0.08$ (IoU $+0.190\pm0.07$) averaged over all combinations and datasets.
\begin{table}[H]
  \centering
  \caption{Comparison of all baselines to ACS after \textbf{30 epochs}.}
    \begin{tabular}{l|l|ll|ll|ll|l|l}
    \multicolumn{1}{l}{} & \multicolumn{1}{l}{} & \multicolumn{2}{l|}{Dataset A} & \multicolumn{2}{l|}{Dataset B} & \multicolumn{2}{l|}{Dataset C} & \multicolumn{2}{l}{\textbf{Average}} \\
\cline{3-10}    \multicolumn{1}{l}{} & \multicolumn{1}{l}{} & IoU   & Dice  & IoU   & Dice  & IoU   & Dice  & \multicolumn{1}{l}{IoU} & Dice \\
    \hline
    \hline
    AB  & Baselines & 0.641 & 0.779 & 0.779 & 0.875 & 0.358 & 0.512 & 0.593 & 0.722 \\
          & \textbf{ACS} (ours) & \textbf{0.749} & \textbf{0.855} & \textbf{0.793} & \textbf{0.884} & \textbf{0.478} & \textbf{0.628} & \textbf{0.673} & \textbf{0.789} \\
    \hline
    \hline
    AC  & Baselines & 0.080 & 0.147 & 0.265 & 0.416 & 0.380 & 0.547 & 0.241 & 0.370 \\
          & \textbf{ACS} (ours) & \textbf{0.646} & \textbf{0.782} & \textbf{0.731} & \textbf{0.844} & \textbf{0.727} & \textbf{0.841} & \textbf{0.702} & \textbf{0.822} \\
    \hline
    \hline
    BC  & Baselines & \textbf{0.260} & \textbf{0.407} & 0.749 & 0.856 & 0.649 & 0.784 & 0.553 & 0.682 \\
          & \textbf{ACS} (ours) & 0.239 & 0.376 & \textbf{0.798} & \textbf{0.887} & \textbf{0.710} & \textbf{0.829} & \textbf{0.582} & \textbf{0.698} \\
    \end{tabular}%
  \label{tab:results_30}%
\end{table}%

\textbf{Stage 2:} To measure overall continual learning performance, i.e., the combination of learning and forgetting, we inspect the average scores over all datasets after 60 epochs in Tab. \ref{tab:results_60}. While the comparison methods' results fluctuate, our approach achieves a consistently higher performance across all combinations and datasets.
This observation manifests in an increase of the average Dice score by $+0.127\pm0.10$ over the U-Net, $+0.156\pm0.03$ over the U-Net-b, $+0.118\pm0.08$ over BS-MAS, and $+0.152\pm0.14$ over OL-KD.
On combination \textit{AB-C}, the U-Net drops by an IoU of $>47\%$ (Dice $>35\%$) on dataset A and by $>26\%$ (Dice $>17\%$) on dataset B. The remaining methods, including ACS, show a significantly lower decline and effectively reduce catastrophic forgetting.
\begin{table}[ht]
  \centering
  \caption{Comparison of all baselines and ACS after \textbf{60 epochs}. ABC represents joint training on all datasets at once.\\}
    \begin{tabular}{l|l|ll|ll|ll|ll}
    \multicolumn{1}{l}{} & \multicolumn{1}{l}{} & \multicolumn{2}{l|}{Dataset A} & \multicolumn{2}{l|}{Dataset B} & \multicolumn{2}{l|}{Dataset C} & \multicolumn{2}{l}{\textbf{Average}} \\
\cline{3-10}    \multicolumn{1}{l}{} & \multicolumn{1}{l}{} & IoU   & Dice  & IoU   & Dice  & IoU   & Dice  & IoU   & Dice \\
    \hline
    \hline
    AB-C  & U-Net & 0.238 & 0.381 & 0.501 & 0.664 & 0.621 & 0.764 & 0.454 & 0.603 \\
          & U-Net-b & 0.339 & 0.503 & 0.570 & 0.724 & 0.473 & 0.631 & 0.461 & 0.619 \\
          & BS-MAS & 0.304 & 0.464 & 0.568 & 0.722 & \textbf{0.624} & \textbf{0.766} & 0.499 & 0.651 \\
          & OL-KD & 0.578 & 0.729 & 0.727 & 0.841 & 0.473 & 0.633 & 0.593 & 0.734 \\
          & \textbf{ACS} (ours) & \textbf{0.640} & \textbf{0.779} & \textbf{0.760} & \textbf{0.863} & 0.572 & 0.718 & \textbf{0.657} & \textbf{0.787} \\
    \hline
    \hline
    AC-B  & U-Net & 0.295 & 0.451 & 0.718 & 0.836 & 0.387 & 0.547 & 0.467 & 0.611 \\
          & U-Net-b & 0.273 & 0.425 & 0.567 & 0.723 & 0.419 & 0.584 & 0.419 & 0.577 \\
          & BS-MAS & 0.307 & 0.466 & 0.702 & 0.825 & 0.364 & 0.523 & 0.458 & 0.604 \\
          & OL-KD & 0.094 & 0.171 & 0.381 & 0.549 & 0.400 & 0.571 & 0.292 & 0.430 \\
          & \textbf{ACS} (ours) & \textbf{0.681} & \textbf{0.808} & \textbf{0.787} & \textbf{0.880} & \textbf{0.679} & \textbf{0.808} & \textbf{0.716} & \textbf{0.832} \\
    \hline
    \hline
    BC-A  & U-Net & \textbf{0.745} & \textbf{0.852} & 0.668 & 0.800 & 0.418 & 0.579 & \textbf{0.610} & \textbf{0.743} \\
          & U-Net-b & 0.450 & 0.615 & 0.591 & 0.742 & 0.497 & 0.661 & 0.513 & 0.673 \\
          & BS-MAS & 0.731 & 0.847 & 0.626 & 0.768 & 0.409 & 0.569 & 0.589 & 0.728 \\
          & OL-KD & 0.347 & 0.511 & \textbf{0.766} & \textbf{0.867} & \textbf{0.639} & \textbf{0.776} & 0.584 & 0.718 \\
          & \textbf{ACS} (ours) & 0.600 & 0.747 & 0.649 & 0.786 & 0.465 & 0.631 & 0.571 & 0.721 \\
    \hline
    \hline
    ABC   & U-Net & 0.277 & 0.431 & 0.458 & 0.627 & 0.431 & 0.599 & 0.388 & 0.440 \\
    (joint) & \textbf{ACS} (ours) & \textbf{0.737} & \textbf{0.847} & \textbf{0.760} & \textbf{0.863} & \textbf{0.724} & \textbf{0.839} & \textbf{0.740} & \textbf{0.849} \\
    \end{tabular}%
  \label{tab:results_60}%
\end{table}%

Combination \textit{AC-B} shows the clear advantage of our approach. Dataset A contains four types of disorders recorded by a single scanner, while dataset C holds three disease patterns recorded by three different scanners. The baselines struggle with the diversity of these domains, and our model outperforms them by an IoU of $>91\%$ (Dice $>53\%$).
These observations show that our model learns a sufficient content representation that can deal with diverse cognitive impairments and scans acquired by scanners of various vendors.

We trace back the low performance on dataset A in combination \textit{BC-A} to A outnumbering B and C in its variability and number of subjects. The high performance of the U-Net thereby originates from overfitting on A which through its high variability still allows it to perform well on B and C. Because ACS is only fine-tuned on A, it cannot fully exploit this anomaly, but still shows competitive results.

\textbf{Ablation Study:} The conducted ablation study in Tab. \ref{tab:ablation} verifies that all losses contribute to the performance of ACS. Only on \textit{AC-B}, the combination of all losses underperforms slightly, but remains competitive. For more detailed numbers we direct the reader to the supplementary material.
\begin{table}[ht]
  \centering
  \caption{Snapshot of the ablation study on ACS trained for 60 epochs with losses active ($\checkmark$) or deactivated ($\text{\sffamily X}$) during training. Average reported over all test datasets in a configuration.}
    \begin{tabular}{llll|ll|ll|ll|ll}
    \multicolumn{1}{l}{} & \multicolumn{1}{l}{} & \multicolumn{1}{l}{} & \multicolumn{1}{l}{} & AB-C  &       & AC-B  &       & BC-A  &       & \multicolumn{2}{l}{\textbf{Average}} \\
    \hline
    \multicolumn{1}{l}{$\mathcal{L}_{adv}^c\:\:\:\:$} & \multicolumn{1}{l}{$\mathcal{L}_{VAE}$} & \multicolumn{1}{l}{$\mathcal{L}_{GAN}$} & \multicolumn{1}{l|}{$\mathcal{L}_{lr}$} & IoU   & Dice  & IoU   & Dice  & IoU   & Dice & IoU & Dice \\
    \hline
    \hline
    $\text{\sffamily X}$ & $\checkmark$ & $\checkmark$ & $\checkmark$ & 0.620  & 0.759 & 0.730 & 0.843 & 0.526 & 0.686 & 0.626 & 0.763 \\
    $\checkmark$ & $\text{\sffamily X}$ & $\text{\sffamily X}$ & $\text{\sffamily X}$ & 0.574 & 0.721 & 0.709 & \textbf{0.849} & 0.524 & 0.681 & 0.603 & 0.750 \\
    $\checkmark$ & $\checkmark$ & $\text{\sffamily X}$ & $\checkmark$ & 0.637 & 0.772 & \textbf{0.732} & 0.843 & 0.555 & 0.711 & 0.642 & 0.776 \\
    $\checkmark$ & $\text{\sffamily X}$ & $\checkmark$ & $\checkmark$ & 0.636 & 0.772 & 0.721 & 0.835 & 0.551 & 0.707 & 0.636 & 0.772 \\
    $\checkmark$ & $\checkmark$ & $\checkmark$ & $\checkmark$ & \textbf{0.658} & \textbf{0.787} & 0.716 & 0.832 & \textbf{0.572} & \textbf{0.721} & \textbf{0.649} & \textbf{0.780} \\
    \end{tabular}%
  \label{tab:ablation}%
\end{table}%

The results demonstrate that leveraging the availability of multiple datasets increases multi-domain segmentation performance by sufficiently learning a domain-invariant representation. This assumption is further supported by the joint training results in Tab. \ref{tab:results_60} showing the superior capability of ACS in comparison to the U-Net. Additionally, our method outperforms the state-of-the-art on most continual learning setups and effectively reduces catastrophic forgetting.

\section{Conclusion}
We propose ACS, an architecture for continual semantic segmentation of multi-domain data that leverages the simultaneous availability of datasets.
In real clinical practice, multiple datasets are available at the beginning of the continual training process through, among other sources, public or accessible historical data. Unlike current methods, we leverage this serendipity to disentangle MRI images into content and domain representations through adversarial training. We then perform multi-domain hippocampal segmentation directly on the domain-invariant content representation.
We demonstrate drastic improvements through domain disentanglement of multi-domain data in the first training stage.
In the second training stage, the benefits of our proposal for continual learning become clear by showcasing that using all available data reduces catastrophic forgetting and outperforms current state-of-the-art methods.
Our method pushes continual learning closer towards a clinical application where various degrees of variability such as disease patterns, scan vendors, and acquisition protocols exist and further enables the continual usage of deep learning models in clinical practice.


\newpage

\bibliographystyle{splncs04.bst}
\bibliography{miccai}

\end{document}